\documentclass[%
 reprint,
 amsmath,amssymb,
 aps,
]{revtex4-2}

\usepackage{graphicx}
\usepackage{dcolumn}
\usepackage{bm}
\usepackage[normalem]{ulem}

\usepackage{xcolor}

\begin{document}

\newcommand{\cred}[1]{\textcolor{black}{#1}}
\newcommand{\credd}[1]{\textcolor{black}{#1}}
\newcommand{\cblue}[1]{\textcolor{black}{#1}}
\newcommand{\cbrown}[1]{\textcolor{black}{#1}}
\newcommand{\cpurp}[1]{\textcolor{black}{#1}}
\newcommand{\bs}{\bar{s}}
\newcommand{\bkappa}{\bar{\kappa}}

\title{Contact geometry in peeling a thin solid from a liquid}
\title{Torque-free peeling: lifting a thin partially-wetting solid off a liquid surface} 
\title{
Torque-free peeling:
\cblue{
the triple-phase contact line at a bendable sheet}}
\title{\cblue{Peeling without local torque: how a bendable sheet detaches from a liquid}}
\title{\cblue{Peeling from a liquid}}

\author{Deepak Kumar$^1$}
\author{Nuoya Zhou${^2}$}
\author{\cblue{Fabian Brau$^{3}$}}
\author{Narayanan Menon${^2}$}%
\author{Benny Davidovitch${^2}$}%
\affiliation{%
 $^1$ Department of Physics, Indian Insitute of Technology Delhi, New Delhi 110016; \\ 
$^2$ Department of Physics, University of Massachusetts Amherst, Amherst MA 01003; \\ 
 \cblue{$^3$ Nonlinear Physical Chemistry Unit, Universite libre de Bruxelles (ULB), CP231, 1050 Brussels, Belgium}}





\date{\today}

\begin{abstract}
We establish the existence of a cusp in the curvature of a solid sheet at its contact with a liquid subphase.  
\cblue{We study two configurations in floating sheets where the solid-vapor-liquid contact line is a straight line and a circle, respectively.} In the former case, a rectangular sheet is lifted at its edge, whereas in the latter a gas bubble is injected beneath \cblue{a floating} sheet. We show that in both geometries the derivative of the sheet’s curvature is discontinuous. 
\cblue{We demonstrate that the boundary condition at the contact is identical in these two geometries, even though the shape of the contact line and the stress distribution in the sheet are sharply different}.

\end{abstract}

\maketitle



The peel test is an extensively used method to measure the strength of adhesion of a \cbrown{sheet} to a substrate. The test, \credd{schematically depicted in Fig.~\ref{fig1} panels B1-D1,} is usually based on measuring the force required to separate the \cbrown{sheet} from the substrate. 
\cblue{In fact,} a direct measurement of the shape of the \cbrown{sheet} near the separation front, \cred{performed in a classic 1930 study  
by Obreimoff on a freshly cleaved mica \cite{obreimoff},}    
\cred{has been arguably the earliest attempt} 
to \cblue{determine} the 
surface energy of solids. 
\cred{In this test}
the \cred{peeled-off} part of the \cbrown{sheet} 
has a parabolic shape while the adhered portion of the \cbrown{sheet} remains flat.  
\cred{The 
discontinuity in curvature, $\kappa(s)$, where $s$ is the distance to the contact line
, reflects a highly-localized torque (on the scale of the \cbrown{sheet}'s thickness) that is exerted by the rigid substrate on the peeled-off \cbrown{sheet}. 
The presence of  a singular torque underlies 
the celebrated Obreimoff's law: 
\begin{gather}
    \text{peeling off rigid substrate:}  \nonumber \\
    [[\kappa]] = \sqrt{2}/\ell_{bc} 
        \label{eq:solid-torque} \\
   \text{where:} \ \  \ell_{bc} \equiv \sqrt{B/T} \label{eq:ellbc}  \ . 
\end{gather}   
\begin{figure}
    \includegraphics[width=0.5\textwidth]{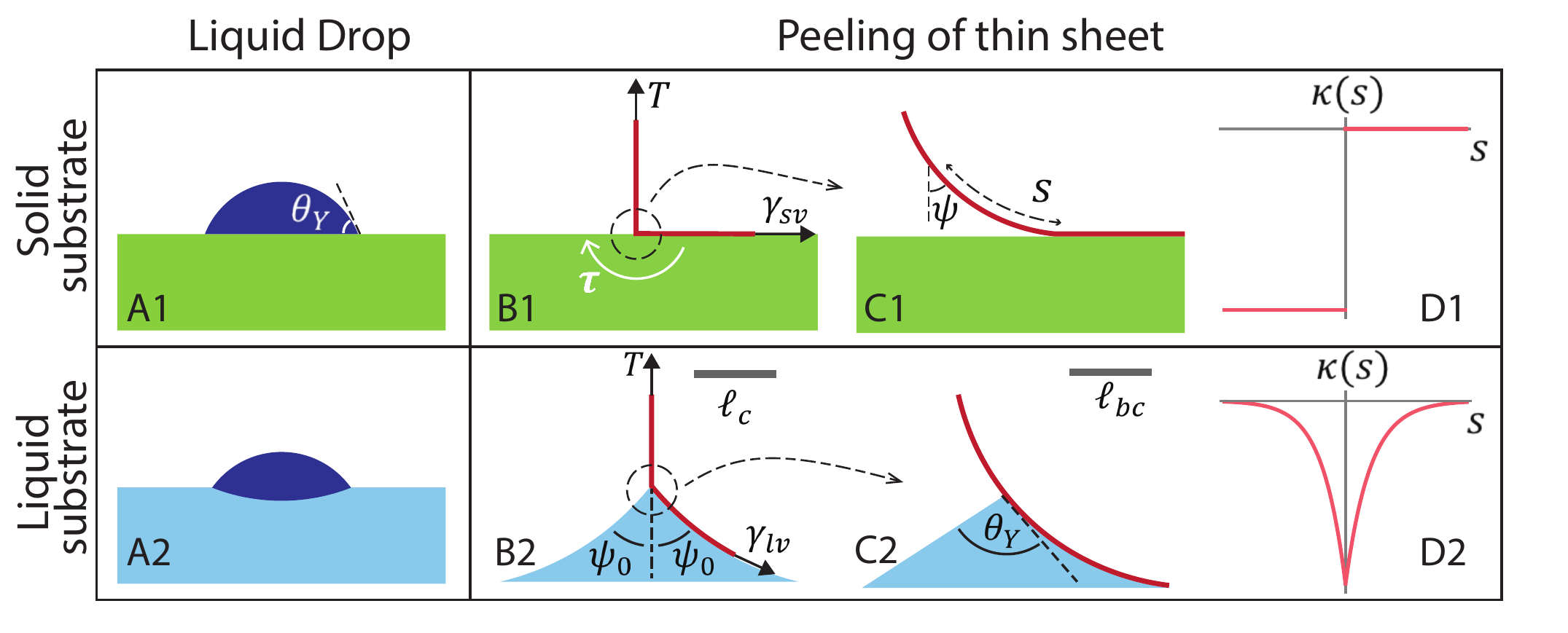}
    \caption{ (A1) Liquid drop on a rigid solid substrate. (A2) Liquid drop on a liquid substrate. (B1, C1) Thin sheet peeling from a solid substrate. (B2, C2) Thin sheet peeling from a liquid substrate. Panels \cblue {B1 and B2 are at the capillary length scale, $\ell_c$, while  panels C1 and C2 are zoomed in to the \cpurp{bendo-}capillary scale $\ell_{bc}$.} (D1) Curvature near the contact line for thin \credd{sheet} peeling from a solid substrate \cite{obreimoff}. (D2) \cpurp{Continuous curvature for a thin \credd{sheet} peeled from a \emph{liquid} substrate characterized by a cusp at the contact line.} } 
    \label{fig1}
\end{figure}
Here, $[[\kappa]]=\kappa(s\to 0^{+})-\kappa(s\to 0^{-})$ denotes the jump in curvature at $s=0$ in the (otherwise continuous) curvature, 
$B$  the bending modulus of the \cbrown{sheet}, and $T$ the tension in the \cbrown{sheet} (attributed by Obreimoff to its surface energy with the ambient phase). In Eq.~(\ref{eq:solid-torque})\cpurp{,} we follow a terminology used in studies of elasto-capillary phenomena that involve slender bodies at fluid interfaces, where Eq.~(\ref{eq:ellbc}) defines  
a ``bendo-capillary'' length, $\ell_{bc}$, over which bending and tensile forces are comparable.  
}

In this 
\cred{paper} we
\cred{study 
peeling of a thin solid \cbrown{sheet} from a \textit{liquid} subphase,} \credd{schematically depicted in Fig.~\ref{fig1} panels B2-D2.} 
In contrast to a \cred{rigid} solid, 
no localized torque is exerted by the liquid subphase at the contact line. 
{
Consequently, the curvature is 
continuous, 
and the only discontinuity possible at mechanical equilibrium is of the 
derivative $\kappa'(s)$. The analogue of Obreimoff's law for peeling a solid \cbrown{sheet} off a liquid subphase becomes: 
\begin{gather}
    \text{peeling off liquid subphase:}  \nonumber \\
    [[\kappa]] = 0 \ \ ; \ \ [[\kappa']]  = \sin\theta_Y / \ell_{bc}^2\ , 
    \label{eq:liquid-no-torque}
\end{gather} 
where $\ell_{bc}$ is defined through Eq.~(\ref{eq:ellbc}) with $T = \gamma_{\ell v}$, and 
\begin{equation}
\theta_Y = \cos^{-1} \cpurp{\left[{(\gamma_{sv}-\gamma_{s\ell})}/{\gamma_{\ell v}}\right]}   
\label{eq:YLD}
\end{equation}
is the Young-Laplace-Dupr\'e (YLD) angle, which is determined by the mutual surface energies between the liquid, solid, and ambient (vapor) phase. While Eq.~(\ref{eq:liquid-no-torque}) has been noted already in a one-dimensional (1D) model system of  
\cpurp{``}bendable'' partial wetting phenomena \cite{Neukirch2013}, 
whereby a finite liquid volume 
is deformed upon making contact with a thin solid along a straight line \cite{Py2007}, our study is the first, to our knowledge, to confirm it experimentally  and employ it in realistic peeling geometries in 1D as well as in 2D (\textit{i.e.} circular contact line). \credd{Determining}
a discontinuity in the derivative of the curvature (which amounts to the third derivative of a profile extracted from an image) is challenging, as noise-averaging smooths over the crucial localized feature we seek to identify.  {\cblue{
Indeed, we are not aware of any direct experimental study 
of a discontinuity in second 
derivative in the solid-peeling case.}}
}  



\cred{As illustrated in Fig.~\ref{fig1}, 
the difference between the original Obreimoff law for peeling off a rigid substrate (Eq.~\cpurp{(}\ref{eq:solid-torque}\cpurp{)}) and its modified version for a  
liquid bath (Eq.~\cpurp{(}\ref{eq:liquid-no-torque}\cpurp{)}), 
parallels the difference between the       
contact angle laws for 
solid-liquid-vapor 
(YLD) and a liquid-liquid-vapor (Neuman). 
 In both scenarios -- partial wetting of a finite liquid volume (panels A) \textit{versus} peeling off a substrate (panels B-D) -- the difference between a rigid solid substrate (top row) and a liquid subphase (bottom row) stems from the fact that a liquid bath cannot support normal load without deforming its surface.  
However, in contrast to the contact angle problem on either a liquid or solid subphase, where the only length scale is the size of the liquid drop, the geometry of a    
sheet  peeled-off from a liquid sub-phase 
consists of multiple scales.  
Zooming in close to the contact line at a size scale $\ll \ell_{bc}$ 
(panel \credd{C}2) reveals a geometry that is 
\credd{almost} indistinguishable from the vicinity of a contact on a thick rigid body 
of the same material (panel A1), 
\credd{except for a 
discontinuity of the 
$3^{rd}$ derivative of the surface, which is reflected (panel D2) by a cusp in the curvature 
$\kappa(s)$.} 
\credd{Z}ooming out to a size that is $ \gg \ell_{bc}$ yet $\ll \ell_c =\sqrt{\gamma_{\ell v}/\rho g}$ (panel B2) one observes a liquid meniscus \credd{dominated} 
by a balance of surface tension and gravity (Young-Laplace equation), 
\credd{which} terminates at a kink, as if the curvature was diverging.}
This multi-scale \credd{scenario} 
is valid only if the \cbrown{sheet} is sufficiently thin, such that 
$\ell_{bc} \ll \ell_c$, or more generally:  
\begin{equation}
    \epsilon \ll 1 \ \ , \ \ \text{where} \ \ \epsilon \equiv (\ell_{bc}/L_{out})^2 \ .
       \label{eq:softness}
\end{equation}
%
Here, $L_{out}$ is an ``outer" scale, at which the curvature approaches an asymptotic value, which is $\ll \ell_{bc}^{-1}$. For the example depicted in Fig.~\ref{fig1}\cpurp{D2} , where $L_{out} = \ell_c$, the ratio $\epsilon = \cpurp{\rho g B/\gamma_{\ell v}^2}$ is akin to the ``softness'' parameter that was defined in Ref.~\cite{Huang2010}. \\
 
\begin{figure}
    \includegraphics[width=0.5\textwidth]{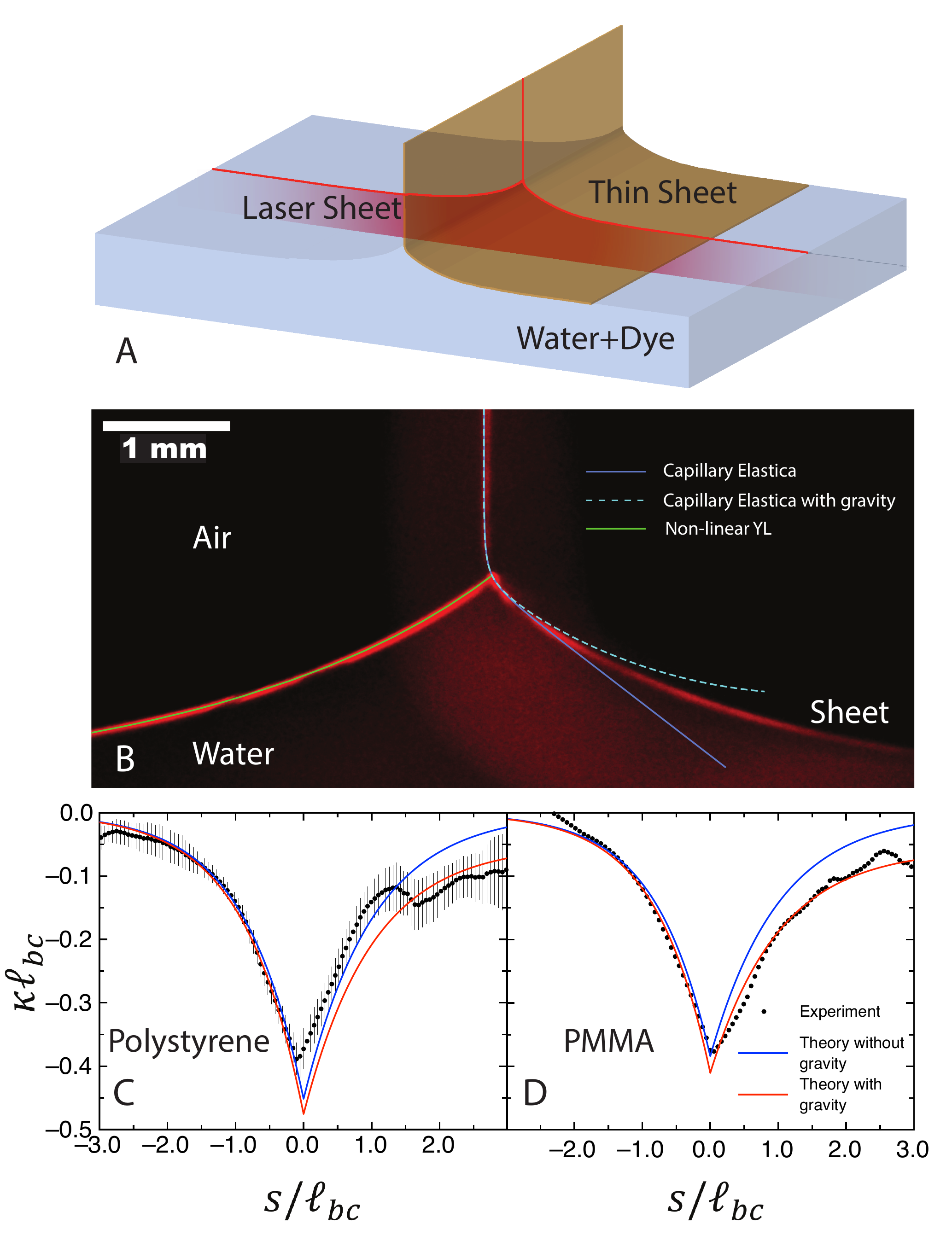}
    \caption{1D geometry. (A) Schematic of experimental setup. The upper edge of the \cbrown{sheet} is held at a fixed, controllable height. (B) A typical image of the sheet profile (Polystyrene, thickness $t=2$ $\mu$\cpurp{m} corresponding to $\ell_{bc}=0.2$ \cpurp{mm}). Superimposed on the image are solutions of the \cbrown{non-linear} Young-Laplace equation \cbrown{\cite{Anderson2006}} (solid green line), the capillary \textit{elastica} \cbrown{Eq.~\cpurp{(\ref{eq:oreder-0})}} (solid blue curve), and the capillary \textit{elastica} with gravity correction (dashed cyan curve). \credd{ The curvature} $\kappa(s)$ for (C) Polystyrene \cbrown{sheet} ($t=2$ $\mu$\cpurp{m}) with estimated statistical error bars and (D) PMMA \cbrown{sheet} ($t=2$ $ \mu$\cpurp{m}). Filled black circles, blue and red lines represent data, theoretical prediction from capillary \textit{elastica} and capillary \textit{elastica} with gravity correction, respectively.}
    \label{fig2}
\end{figure}
\paragraph*{1D translationally symmetric geometry:}
We first address  
an effectively one-dimensional (1D) geometry, where the deformed sheet is characterized by translational symmetry along the direction parallel to the contact line,     
as shown schematically in the bottom part of Fig.~\ref{fig1}B and \cpurp{Fig.~\ref{fig2}A}. 
Such a 1D set-up is realized in   
a floating, rectangular thin \cbrown{sheet}, which is peeled off by exerting a vertical force $T_{peel}$ along one of its short edges. As we noted above, 
when observed at intermediate scales, 
$\ell_{bc} \ll |s| \lesssim \ell_c$, the
\cbrown{sheet} appears to have a cusp at the contact line; furthermore,  
the mechanical equilibrium shape is 
characterized by reflection symmetry of the two sides of the surface (\cbrown{sheet}-covered and liquid-vapor) around the vertical line
\cite{Kumar2020}.
The reflection symmetry indicates that the tension in the wet part of the \cbrown{sheet} is identical to the liquid-vapor surface tension, $\gamma_{\ell v}$, 
and force balance at the contact line thus determines the force, $T_{peel}$, the opening angle, $2\psi_{0}$, of the apparent cusp, and the height, $H_0$, of the contact line 
over the liquid bath level: 
\begin{gather}
T_{peel} = \gamma_{\ell v}(1+ \cos\theta_Y)  = 2\gamma_{\ell v} \cos \psi_{0}
\label{eq:forcebalance} \\
\Longrightarrow  \ \psi_{0} \approx  \cos^{-1}[\cos^2\frac{\theta_Y}{2}] \  \ ,
\ \ H_0 \approx f(\psi_0) \ell_c  \ ,   
\label{eq:psiinf}
\end{gather}
where the function $f(\psi_0)$ is found by solving the (nonlinear) Young-Laplace equation \cite{Anderson2006}, such that $f\!\approx\! (\pi/2-\psi_0)$ for $\psi_0 \to \pi/2$ and $f\!\approx \!\sqrt{2}$ for $\psi_0 \to  0$.
The shape of the whole \cbrown{sheet} is described by a planar vector, $\vec{X}(s) \!= \!x(s)\hat{i} \!+\! y(s)\hat{j}$, where $s$ is an arclength parameter, and is 
conveniently described through \credd{the}    
angle, $\psi(s)$, between the tangent vector, $\vec{X}'$ and the downward vertical $-\hat{y}$:  
\begin{equation}
x'(s) = \sin\psi \ \ ; \ \ y'(s) = \cos\psi  \ \ ;  \ \ \text{and} \ \ \kappa(s) = \psi'     \ . 
\label{eq:basic}
\end{equation}
At mechanical equilibrium, 
the shape satisfies 
the  
capillary {\emph{ elastica}}, which expresses normal force balance 
\cite{Neukirch2013,kozyreff2022wetting}:  
\begin{gather}
B \left[\kappa'' + \frac{\kappa^3}{2}\right] - T(s) \kappa  = - P(s)
\cpurp{+} \gamma_{\ell v} \sin\theta_Y \delta_d (s)   \nonumber \\ 
P(s) \!= \!\left\{ \begin{array}{ll}
         0  & s < 0  \\
        -\rho g y(s) & s > 0  \end{array} \right. 
\quad  
T(s) = \left\{ \begin{array}{ll}
         T_{peel} & 
         s < 0\\
        \gamma_{\ell v} & s > 0  \end{array} \right. 
\label{eq:cap-elastica}
\end{gather}
Here, $P(s)$ 
is the hydrostatic pressure exerted by the liquid bath on the wet portion of the \cbrown{sheet}, $s>0$ (where $\rho$ and $g$ are, respectively, the liquid density and gravity acceleration), $T(s)$ is the tension in the \cbrown{sheet}, and 
$\gamma_{\ell v}\sin\theta_Y$ is the normal force exerted by the liquid-vapor interface at the contact line. In a 1D geometry at mechanical equilibrium (and absence of external shear forces) the tension $T(s)$ satisfies $\partial_s T = 0$, and is consequently constant in the dry part ($s<0$), where it is given by the force $T_{peel}$ exerted by the peeler, and in the wet part ($s>0$), where it is given by the liquid-vapor surface that pulls on the edge of the floating \cbrown{sheet}.   

Integrating both sides of Eq.~
(\ref{eq:cap-elastica}) over an infinitesimal neighborhood of the contact line, $s=0$, 
we obtain: 
\begin{equation}
\label{eq:findpsicon-A} 
B[[\kappa']] = 
\cpurp{\gamma_{\ell v} \sin\theta_Y} \ , 
\end{equation}
and integrating once more across the contact line we obtain 
$[[\kappa]] = 
0$,   
thereby establishing Eq.~(\ref{eq:liquid-no-torque})
\footnote{\credd{
A higher-order effect, which cannot be accounted by Eq.~(\ref{eq:cap-elastica}), it the small torque exerted by the liquid-vapor interface on the \cbrown{sheet} if they are not perpendicular at the contact line (\textit{i.e.} $\theta_Y \neq \pi/2$). This localized torque is explicitly proportional to the thickness, yielding a discontinuity of the curvature $[[\kappa]] \sim \cos(\theta_Y) \cdot t/\ell_{bc}^2$, whose effect on the shape is negligible
, \textit{i.e.} $[[\kappa']] \gg [[\kappa]]/\ell_{bc}$ for $t \!\ll\! \ell_{bc}$.}}.
 %

On each side of the contact line, the profile of the sheet is determined by the capillary \textit{elastica}~\eqref{eq:cap-elastica} which is a nonlinear $3^{rd}$ order equation for $\psi(s)$, whose solution requires 3 boundary conditions (BCs). Thus, in addition to Eqs.~(\ref{eq:findpsicon-A},\ref{eq:liquid-no-torque}), 4 other BCs must be specified. To obtain these, we non-dimensionalize length by defining $\bs = s/\ell_{bc}$ and $\bkappa = \kappa \ell_{bc}$, and consider Eq.~(\ref{eq:cap-elastica}), in the \textit{singular} limit $\epsilon \to 0$ (see SI). At $O(\epsilon^0)$, we obtain 2 ``outer'' BCs at each side of the contact line. At 
$\bs \to -\infty:  \ \psi \to 0, \bkappa \to 0$, and at $\bs \to +\infty:  
\ \psi \to \psi_0, \bkappa \to 0$, and the corresponding (exact) solution of Eq.~(\ref{eq:cap-elastica}) at $O(\epsilon^0)$ is given by:
\begin{eqnarray}
\bs \! &>& \! 0\!: \  \bkappa \! = \! 
-2 \ \text{sech}(\bs \!+\! \bs_w) \  , 
\label{eq:oreder-0}\\ 
\bs \! &<& \! 0\!:\  \bkappa \! = \! 
-2 \sqrt{r} \ \text{sech} 
{[(\bs \!+\! \bs_d}){\sqrt{r}]}  \ ,   r \equiv \frac{T_{peel}}{\gamma_{\ell v}} = 2\cos\psi_0 \  ,
\nonumber 
\end{eqnarray}
\begin{gather}
\text{where:} \  \    
\bs_w = \cosh^{-1}(a)  \  ; 
\ \bs_d =-\cosh^{-1}(a \sqrt{r})   /\sqrt{r} \ ; \ \nonumber \\
   a = (2/\sin\theta_Y)
   \sqrt{b + \sqrt{b^2-1}} \ ; \ b = 1+r \ . 
\end{gather}
In the SI\cpurp{,} we describe a next-order, $O(\epsilon^1)$ solution, which incorporates the gravity effect on the sheet curvature and is 
useful for comparison with experimental data.

The setup employed to study the problem experimentally is illustrated schematically in \cpurp{Fig.}\ref{fig2} A. Thin sheets of Polystyrene (Polymer source P10453-S, \cpurp{$Mw=97.6$ kDa}, Young's modulus = 3.5 GPa) and PMMA (Aldrich 182265, \cpurp{$Mw\sim 996$ kDa}, Young's modulus = 3 GPa) of 1-2 $\mu m$ thickness are prepared by spin-coating on glass slides. Rectangles of \cpurp{20 mm} width and \cpurp{60 mm} length are cut from these films and floated on the surface of a water bath ($\ell_c$=\cpurp{2.7 mm}). The long end of the film is then lifted out of the water surface by using a triangular hanger made up of graphite rods of diameter \cpurp{0.7 mm} (pencil leads). Sheet illumination produced from a green laser (wavelength \cpurp{532 nm}) is used to illuminate the interface near the contact line for imaging. A dye (Rhodamine-B) is dissolved in water in miniscule amount rendering both sides of the interface fluorescent. The interface is imaged using a DSLR camera (Nikon D5300) with a macro-lens and a long pass filter to admit only the fluorescent light. The laser sheet is positioned near the center of the film which is many $\ell_c$ away from the ends of the contact line. In this configuration end-effects near the edges of the sheet are negligible and the film profile can be assumed to be 2D. A typical image obtained from the setup is shown in \cpurp{Fig.} \ref{fig2}B. The resolution of the imaging setup (1 pixel $\sim 1$ $\mu m$) is typically much smaller than $\ell_{bc}$, \credd{which is} approximately \cpurp{0.2 mm} for the films used. Superimposed on the image are the solution of the Young-Laplace equation \cite{Anderson2006} as green solid curve \cblue{for the liquid-vapor interface (left to the contact line)}, the solution to the capillary \textit{elastica} without gravity as a blue solid curve, and the solution of the capillary \textit{elastica} with gravity as the dashed cyan curve.

 A gradient method is used to detect the interface and to obtain its $(x,y)$ coordinates along the film from the images, and $\kappa(s)$, the curvature {\emph{vs.} arclength is computed from:
 \begin{gather}
 \kappa(x)\!=\!\frac{y''(x)}{(1+y'(x)^2)^{3/2}} 
 \ ; \ s(x)\!=\!\int_{0}^{x}\!\!\!\sqrt{1+y'(x)^2} dx \ . 
 \end{gather} 
On computing derivatives from experimental data, the noise in the data gets amplified, which usually
\cblue{necessitates} 
some form of smoothing. However, traditional smoothing methods will suppress any cusp  in $\kappa(s)$. 
We therefore developed the 
algorithm described below to extract $\kappa(s)$ from the $(x,y)$ data.


\cblue{We divide the whole data set into intervals of length $\Delta\sim \ell_{bc}$. We construct a sample of this data by choosing one data point from each interval randomly with a uniform probability distribution.  We can estimate the position of the contact line from the images with a much higher accuracy and precision of a few pixels.  We add to the data sample a contact-line location selected randomly with a Gaussian probability distribution centered at the estimated position of the contact line and having a width equal to the estimated error. A spline function of order $3$ made up of Hermite polynomials is generated using this sampling of data points and the curvature is computed on this spline function at roughly every $10^{th}$ point of the original data set. The process is repeated a large number of times (about twice the number of data points in each interval), selecting a different sampling of data points, such that the whole data set is adequately represented. The curvature profiles obtained from individual data samples are averaged to obtain the final $\kappa(s)$ curve. This procedure allows noise-averaging and use of the full data set without spatial averaging that would smooth the putative curvature cusp. }

The black filled circles in \cpurp{Fig.} \ref{fig2}C-D 
show 
$\kappa(s)$ \credd{as} 
determined by the above described method for a polystyrene (PS) and a PMMA film of thickness $t=2$ $\cpurp{\mu}$m, respectively. Superimposed on the experimental data are the theoretical predictions obtained by solving the capillary \textit{elastica} equation neglecting gravity and capillary \textit{elastica} with gravity in blue and red lines respectively. We notice that the theoretical predictions match the experimental data quite well and show a clear cusp at $s=0$. 

When gravity is neglected, the only input parameter required to solve 
the capillary \textit{elastica} 
is $\psi_0$; however, this can be directly measured from the water-air interface near the contact line in the image and is found to be $48.8 ^\circ$ and $40.0 ^\circ$ for the PS and the PMMA films in \cpurp{Fig.} \ref{fig2}C and \cpurp{\ref{fig2}}D, respectively. In order to generate the solution of the capillary \textit{elastica} with gravity, in addition to $\psi_0$, we require the value of $\epsilon=(\ell_{bc}/\ell_c)^2$, which is already known. Thus, there is no fitting parameter involved in computing the \cpurp{theory} curves. \\

\begin{figure}
    \includegraphics[width=0.5\textwidth]{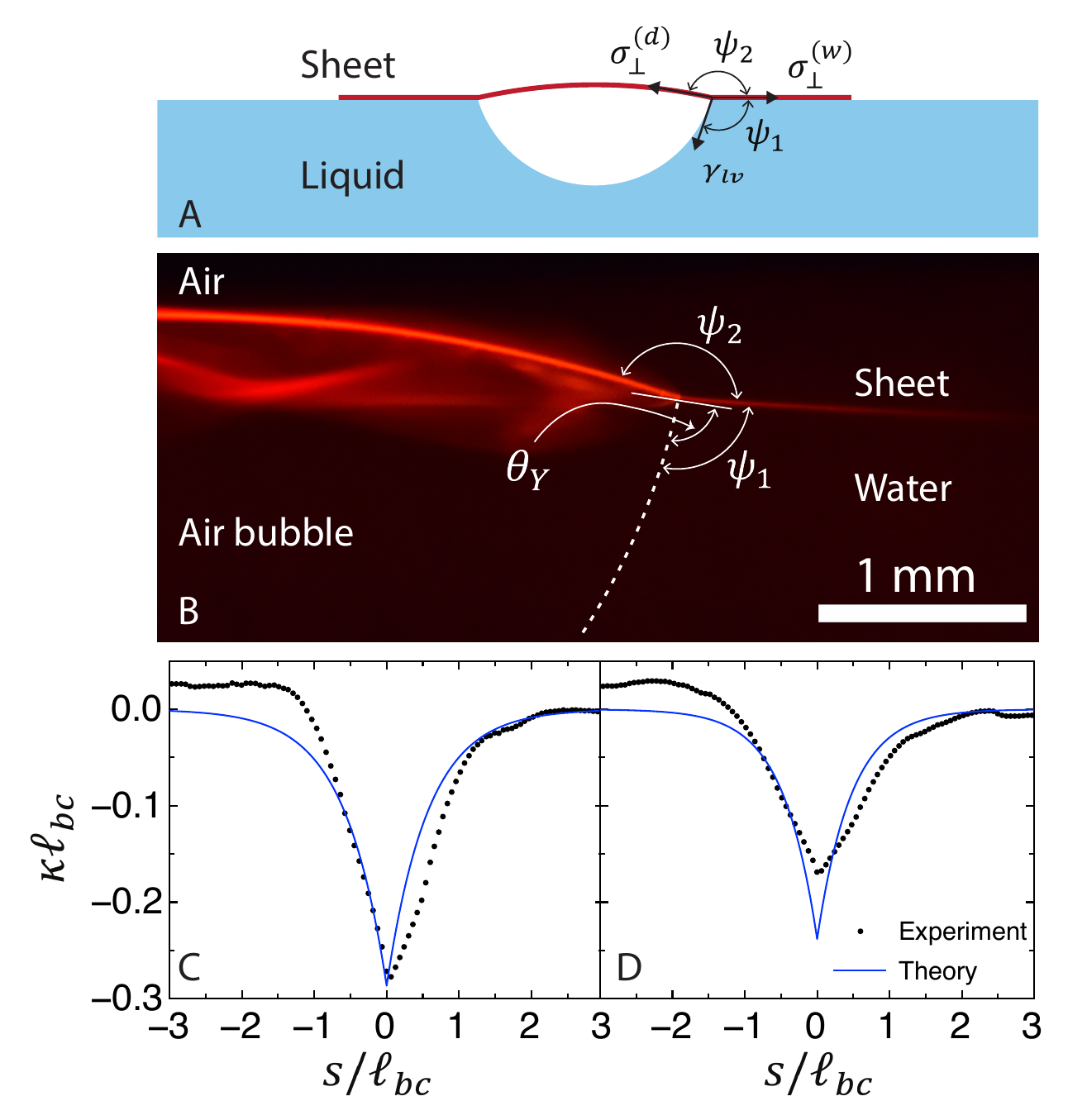}
    \caption{
    2D axially-symmetric geometry. (A) The setup consists of a thin \cbrown{sheet} floating on water with an air bubble underneath. (B) A typical image of the \cbrown{sheet} profile (Polystyrene, thickness $t=2$ $\cpurp{\mu}$m). The dashed white curve is a circle fitted to the air bubble\cbrown{, and the white \cpurp{solid} line is tangent to the \cbrown{sheet} profile at the contact line.} \credd{The curvature} $\kappa(s)$ for air bubble of radius (C) $4.1$ \cpurp{mm} and (D) $3.1$ \cpurp{mm}.}
    \label{fig3}
\end{figure}
\paragraph*{Axial geometry:} 
While the \cblue{1D} geometry of the setup described 
in Fig.~\ref{fig2} presents a simple setting to discuss the boundary conditions at the contact line, the curvature cusp predicted by Eq.~(\ref{eq:liquid-no-torque}) appears in various other \cblue{settings that are often encountered in elasto-capillary phenomena.} 
\cblue{One example is the axially-symmetric geometry of a  
}
thin \cbrown{sheet} floating on water with an air-bubble of volume $V$ underneath it, as shown schematically in Fig. \ref{fig3}A. In contrast to the 1D geometry, which is free of Gaussian curvature and whose mechanical equilibrium is thus described everywhere by a planar curve $\vec{X}(s)$
that solves 
the {\emph{elastica}}, 
Eq.~(\ref{eq:cap-elastica}), the axial geometry in Fig.~\ref{fig2} is characterized by Gaussian curvature, and 
thus involves a nontrivial variation of both radial and hoop components of the stress and curvature tensors with radial distance $r$. Hence, the shape of the \cbrown{sheet} must be described by a surface $\vec{X}(r,\theta)$, that is obtained by solving the {F\"oppl-von K\'arm\'an} (FvK) equations \cite{Schroll2013,Davidovitch2018}, and is furthermore susceptible to radial wrinkling instability due to hoop compression exhibited by an axisymmetric solution \cite{Huang2007,Davidovitch2011}. Nevertheless, as long as
the 
bendo-capillary length is sufficiently small, (namely, $\epsilon \ll 1$, where $L_{out}$ in Eq.~(\ref{eq:softness}) is now given by the drop's radius and/or the capillary length $\ell_c$), 
 the dominant terms in the curvature and stress tensors in the vicinity of the contact line are the radial components, and consequently the normal force balance is given by an equation similar to Eq.~(\ref{eq:cap-elastica}): 
\begin{gather}
B \left(\kappa_{\perp}'' + {\kappa_{\perp}^3}/2\right) -\sigma_{\perp} \kappa_{\perp}  = 
- \gamma_{\ell v} \sin\theta_Y \delta_d(s)   
 \ , 
\label{eq:cap-elastica-1}
\end{gather}
where $\kappa_{\perp}(s), \sigma_{\perp}(s)$ are principal components of the curvature and stress tensors, respectively, along the radial direction, perpendicular to the contact line, and $s$ is the  distance from the contact line. In the preceding analysis we neglected in Eq.~(\ref{eq:cap-elastica-1}) the sub-dominant hydrostatic pressure term; for similar reasons we  ignore spatial variation of $\sigma_{\perp}$ on each side of the contact line. Once again, considering an  infinitesimal vicinity of the contact line, we note that 
tangential force balance yields: 
\begin{equation}
\sigma_{\perp}^{(d)}  = \gamma_{lv} \cos\theta_Y + 
\sigma_{\perp}^{(w)}  \ 
\label{eq:tang-force}
\end{equation}
(where superscripts refer to the dry and wet sides), to which we refer as YLD equation, 
and integration of Eq.~(\ref{eq:cap-elastica-1}) 
yields the jump condition, Eq.~(\ref{eq:liquid-no-torque}). 

The validity of Eq.~(\ref{eq:cap-elastica-1}) hinges upon scale separation, namely $\epsilon \ll 1$, where the ratio $\epsilon$ is given by 
Eq.~(\ref{eq:softness}), with  
\begin{equation} \ell_{bc} \longrightarrow  \ell_{bc}^* \equiv \sqrt{B/\sigma_{\perp}} 
\ \ ; \ \ L_{out} \longrightarrow \min\{\ell_C, V^{1/3} \}
\end{equation} 
(see SI and Ref.~\cite{Davidovitch2018}). Similarly to our analysis of Eq.~(\ref{eq:cap-elastica}), the $O(\epsilon^0)$ boundary conditions 
for Eq.~(\ref{eq:cap-elastica-1}) 
consist of vanishing curvature away from the contact line (\textit{i.e.} $\bkappa, \bkappa^{'} \sim O(\epsilon,\sqrt{\epsilon})$, respectively). 
However, in contrast to the simpler 1D geometry, finding the asymptotic tangent angle $\psi_{1,2}$ at the two sides of the contact line, as well as the stress $\sigma_{\perp}$ in its vicinity, requires one to solve the FvK equations -- a nonlinear set of partial differential equations -- in the singular limit of vanishing bending rigidity (known as ``membrane limit'' or ``tension field theory").   
Rather than following such a theoretical track 
(\'a la Refs.~\cite{Schroll2013,Davidovitch2018}), 
we note that force balance on an ``intermediate box'' of size $\ell_{bc}^* \ll \ell \ll L_{out}$, around the contact line (see Fig.~\ref{fig3}B). 
implies that, at $O(\epsilon^0)$:
\begin{gather}
\sigma_{\perp}^{(d)}  = - \sigma_{\perp}^{(w)} \cos\psi_2 - \gamma_{lv} \cdot \cos(\psi_1+\psi_2) \nonumber  \\
\sigma_{\perp}^{(w)} \sin\psi_2  = - \gamma_{lv} \sin (\psi_1+\psi_2) 
\label{eq:Neuman}
 \end{gather} 
 (often called ``Neuman contact'' \cite{Schulman2015}), which implies that $\gamma_{lv}$ and the two asymptotic angles, $\psi_1,\psi_2$, uniquely determine the in-plane stress \cbrown{in the sheet}, $\sigma_{\perp}^{(w)}, \sigma_{\perp}^{(d)}$, near the contact line, and consequently the YLD angle $\theta_Y$ by 
 Eq.~(\ref{eq:tang-force}). Note that for the 1D peeling geometry 
 considered earlier, $\sigma_{\perp}^{(w)} = \gamma_{lv}$, whereas $\psi_1=2\psi_0$ and $\psi_2 = \pi - \psi_0$, such that Eqs.~(\ref{eq:tang-force}, \ref{eq:Neuman}) reduce to Eq.~(\ref{eq:forcebalance}).  \\

An air bubble released \cpurp{ within} the fluid, forms a bubble \cpurp{beneath} the sheet.  To image the contact line between bubble, sheet and water subphase, a vertical plane passing through the center of the setup is imaged using a laser-sheet fluorescence method similar to the one illustrated in Fig. \ref{fig2}. \cpurp{A} typical image of the \cbrown{sheet} profile obtained \cpurp{from} these experiments \cpurp{is shown} in Fig. \ref{fig3}B. A bright-field image \cpurp{is} taken after \cpurp{the fluorescence image}, and used to obtain the profile of the air-bubble. The dashed white curve in Fig. \ref{fig3}B represents a circle fitted to the air-bubble shape. Figures \ref{fig3}C-D 
show $\kappa(s)$ for bubble radii $R=4.1$ \cpurp{mm} and $R=3.1$ \cpurp{mm}, respectively. The data demonstrate that in this geometry too,  the curvature has a cusp near $s=0$, representing a discontinuity in the derivative of the curvature. \\

\credd{
Problems such as the 1D and axial peeling geometries in highly-bendable sheets typically come in two parts with a big separation in length scale - an ``inner'', bending-dominated region of size $\ell_{bc}^*$ that is governed by the \textit{elastica}, and an ``outer'' region, where the shape and stress are independent of bending rigidity.   
At the innermost part of the bending-dominated 
zone is the purely local effect that we have established in this article, with a discontinuity in the gradient of the curvature in the vicinity of the contact line, $[[\kappa_{\perp}']]$, which is determined purely by material parameters ($B, \gamma_{lv},\theta_Y$). This discontinuity affects the bending-dominated region as a ``near-field"
 boundary condition to the \textit{elastica} problem, but a complete solution of the  \textit{elastica} requires also a ``far-field" boundary condition, which is obtained by   
matching with the outer, bending-independent problem.   
In the cases we considered, this matching condition 
is expressed through a single  
parameter, 
the asymptotic angle $\psi_0$ in Fig.~1 or equivalently $\psi_2 = \pi -\psi_0$ in Fig.~3. Neglecting the bending-dominated region altogether (as was proposed in  \cite{Twohig2018} for sufficiently thin sheets)  leads to an error in the region close to the contact. Neglecting the       
curvature discontinuity at the contact line (as in the \textit{elastica} problem for 1D delamination studied by \cite{Wagner2011}) can also lead to an  error in the predicted shape, and when tension is small, 
the error may span a large portion of the sheet.}
%
\credd{In conclusion, we note that 
the geometry-independent nature of the discontinuity, $[[\kappa_{\perp}']]$, provides the basis for a robust method for determining contact angles both at and away
from equilibrium.}

\paragraph*{Acknowledgement}
This research was supported in part by \cbrown{IIT Delhi New Faculty Seed Grant (DK), SERB, India under the grant SRG/2019/000949 (DK)}, and the National Science Foundation under grants NSF-DMR 1822439 (BD) and NSF-DMR 190568 (NM and NZ).

\bibliography{Ref}

\begin{thebibliography}{15}%
\makeatletter
\providecommand \@ifxundefined [1]{%
 \@ifx{#1\undefined}
}%
\providecommand \@ifnum [1]{%
 \ifnum #1\expandafter \@firstoftwo
 \else \expandafter \@secondoftwo
 \fi
}%
\providecommand \@ifx [1]{%
 \ifx #1\expandafter \@firstoftwo
 \else \expandafter \@secondoftwo
 \fi
}%
\providecommand \natexlab [1]{#1}%
\providecommand \enquote  [1]{``#1''}%
\providecommand \bibnamefont  [1]{#1}%
\providecommand \bibfnamefont [1]{#1}%
\providecommand \citenamefont [1]{#1}%
\providecommand \href@noop [0]{\@secondoftwo}%
\providecommand \href [0]{\begingroup \@sanitize@url \@href}%
\providecommand \@href[1]{\@@startlink{#1}\@@href}%
\providecommand \@@href[1]{\endgroup#1\@@endlink}%
\providecommand \@sanitize@url [0]{\catcode `\\12\catcode `\$12\catcode
  `\&12\catcode `\#12\catcode `\^12\catcode `\_12\catcode `\%12\relax}%
\providecommand \@@startlink[1]{}%
\providecommand \@@endlink[0]{}%
\providecommand \url  [0]{\begingroup\@sanitize@url \@url }%
\providecommand \@url [1]{\endgroup\@href {#1}{\urlprefix }}%
\providecommand \urlprefix  [0]{URL }%
\providecommand \Eprint [0]{\href }%
\providecommand \doibase [0]{https://doi.org/}%
\providecommand \selectlanguage [0]{\@gobble}%
\providecommand \bibinfo  [0]{\@secondoftwo}%
\providecommand \bibfield  [0]{\@secondoftwo}%
\providecommand \translation [1]{[#1]}%
\providecommand \BibitemOpen [0]{}%
\providecommand \bibitemStop [0]{}%
\providecommand \bibitemNoStop [0]{.\EOS\space}%
\providecommand \EOS [0]{\spacefactor3000\relax}%
\providecommand \BibitemShut  [1]{\csname bibitem#1\endcsname}%
\let\auto@bib@innerbib\@empty
\bibitem [{\citenamefont {Obreimoff}(1930)}]{obreimoff}%
  \BibitemOpen
  \bibfield  {author} {\bibinfo {author} {\bibfnamefont {J.}~\bibnamefont
  {Obreimoff}},\ }\bibfield  {title} {\bibinfo {title} {The splitting strength
  of mica},\ }\href@noop {} {\bibfield  {journal} {\bibinfo  {journal}
  {Proceedings of the Royal Society of London. Series A, Containing Papers of a
  Mathematical and Physical Character}\ }\textbf {\bibinfo {volume} {127}},\
  \bibinfo {pages} {290} (\bibinfo {year} {1930})}\BibitemShut {NoStop}%
\bibitem [{\citenamefont {Neukirch}\ \emph {et~al.}(2013)\citenamefont
  {Neukirch}, \citenamefont {Antkowiak},\ and\ \citenamefont
  {Marigo}}]{Neukirch2013}%
  \BibitemOpen
  \bibfield  {author} {\bibinfo {author} {\bibfnamefont {S.}~\bibnamefont
  {Neukirch}}, \bibinfo {author} {\bibfnamefont {A.}~\bibnamefont
  {Antkowiak}},\ and\ \bibinfo {author} {\bibfnamefont {J.-J.}\ \bibnamefont
  {Marigo}},\ }\bibfield  {title} {\bibinfo {title} {The bending of an elastic
  beam by a liquid drop: a variational approach},\ }\href
  {https://doi.org/10.1098/rspa.2013.0066} {\bibfield  {journal} {\bibinfo
  {journal} {Proceedings of the Royal Society A: Mathematical, Physical and
  Engineering Sciences}\ }\textbf {\bibinfo {volume} {469}},\ \bibinfo {pages}
  {20130066} (\bibinfo {year} {2013})}\BibitemShut {NoStop}%
\bibitem [{\citenamefont {Py}\ \emph {et~al.}(2007)\citenamefont {Py},
  \citenamefont {Reverdy}, \citenamefont {Doppler}, \citenamefont {Bico},
  \citenamefont {Roman},\ and\ \citenamefont {Baroud}}]{Py2007}%
  \BibitemOpen
  \bibfield  {author} {\bibinfo {author} {\bibfnamefont {C.}~\bibnamefont
  {Py}}, \bibinfo {author} {\bibfnamefont {P.}~\bibnamefont {Reverdy}},
  \bibinfo {author} {\bibfnamefont {L.}~\bibnamefont {Doppler}}, \bibinfo
  {author} {\bibfnamefont {J.}~\bibnamefont {Bico}}, \bibinfo {author}
  {\bibfnamefont {B.}~\bibnamefont {Roman}},\ and\ \bibinfo {author}
  {\bibfnamefont {C.~N.}\ \bibnamefont {Baroud}},\ }\bibfield  {title}
  {\bibinfo {title} {Capillary origami: spontaneous wrapping of a droplet with
  an elastic sheet},\ }\href@noop {} {\bibfield  {journal} {\bibinfo  {journal}
  {Physical review letters}\ }\textbf {\bibinfo {volume} {98}},\ \bibinfo
  {pages} {156103} (\bibinfo {year} {2007})}\BibitemShut {NoStop}%
\bibitem [{\citenamefont {Huang}\ \emph {et~al.}(2010)\citenamefont {Huang},
  \citenamefont {Davidovitch}, \citenamefont {Santangelo}, \citenamefont
  {Russell},\ and\ \citenamefont {Menon}}]{Huang2010}%
  \BibitemOpen
  \bibfield  {author} {\bibinfo {author} {\bibfnamefont {J.}~\bibnamefont
  {Huang}}, \bibinfo {author} {\bibfnamefont {B.}~\bibnamefont {Davidovitch}},
  \bibinfo {author} {\bibfnamefont {C.~D.}\ \bibnamefont {Santangelo}},
  \bibinfo {author} {\bibfnamefont {T.~P.}\ \bibnamefont {Russell}},\ and\
  \bibinfo {author} {\bibfnamefont {N.}~\bibnamefont {Menon}},\ }\bibfield
  {title} {\bibinfo {title} {{Smooth cascade of wrinkles at the edge of a
  floating elastic film}},\ }\href
  {https://doi.org/10.1103/PhysRevLett.105.038302} {\bibfield  {journal}
  {\bibinfo  {journal} {Physical Review Letters}\ }\textbf {\bibinfo {volume}
  {105}},\ \bibinfo {pages} {2} (\bibinfo {year} {2010})},\ \Eprint
  {https://arxiv.org/abs/0901.2892} {0901.2892} \BibitemShut {NoStop}%
\bibitem [{\citenamefont {Anderson}\ \emph {et~al.}(2006)\citenamefont
  {Anderson}, \citenamefont {Bassom},\ and\ \citenamefont
  {Fowkes}}]{Anderson2006}%
  \BibitemOpen
  \bibfield  {author} {\bibinfo {author} {\bibfnamefont {M.~L.}\ \bibnamefont
  {Anderson}}, \bibinfo {author} {\bibfnamefont {A.~P.}\ \bibnamefont
  {Bassom}},\ and\ \bibinfo {author} {\bibfnamefont {N.}~\bibnamefont
  {Fowkes}},\ }\bibfield  {title} {\bibinfo {title} {{Exact solutions of the
  Laplace-Young equation}},\ }\href {https://doi.org/10.1098/rspa.2006.1744}
  {\bibfield  {journal} {\bibinfo  {journal} {Proceedings of the Royal Society
  A: Mathematical, Physical and Engineering Sciences}\ }\textbf {\bibinfo
  {volume} {462}},\ \bibinfo {pages} {3645} (\bibinfo {year}
  {2006})}\BibitemShut {NoStop}%
\bibitem [{\citenamefont {Kumar}\ \emph {et~al.}(2020)\citenamefont {Kumar},
  \citenamefont {Russell}, \citenamefont {Davidovitch},\ and\ \citenamefont
  {Menon}}]{Kumar2020}%
  \BibitemOpen
  \bibfield  {author} {\bibinfo {author} {\bibfnamefont {D.}~\bibnamefont
  {Kumar}}, \bibinfo {author} {\bibfnamefont {T.~P.}\ \bibnamefont {Russell}},
  \bibinfo {author} {\bibfnamefont {B.}~\bibnamefont {Davidovitch}},\ and\
  \bibinfo {author} {\bibfnamefont {N.}~\bibnamefont {Menon}},\ }\bibfield
  {title} {\bibinfo {title} {{Stresses in thin sheets at fluid interfaces}},\
  }\href {https://doi.org/10.1038/s41563-020-0640-9} {\bibfield  {journal}
  {\bibinfo  {journal} {Nature Materials}\ }\textbf {\bibinfo {volume} {19}},\
  \bibinfo {pages} {690} (\bibinfo {year} {2020})}\BibitemShut {NoStop}%
\bibitem [{\citenamefont {Kozyreff}\ \emph {et~al.}(2022)\citenamefont
  {Kozyreff}, \citenamefont {Davidovitch}, \citenamefont {Prasath},
  \citenamefont {Palumbo},\ and\ \citenamefont {Brau}}]{kozyreff2022wetting}%
  \BibitemOpen
  \bibfield  {author} {\bibinfo {author} {\bibfnamefont {G.}~\bibnamefont
  {Kozyreff}}, \bibinfo {author} {\bibfnamefont {B.}~\bibnamefont
  {Davidovitch}}, \bibinfo {author} {\bibfnamefont {S.~G.}\ \bibnamefont
  {Prasath}}, \bibinfo {author} {\bibfnamefont {G.}~\bibnamefont {Palumbo}},\
  and\ \bibinfo {author} {\bibfnamefont {F.}~\bibnamefont {Brau}},\ }\href@noop
  {} {\bibinfo {title} {Wetting of an elastic sheet subject to external
  tension}} (\bibinfo {year} {2022}),\ \Eprint
  {https://arxiv.org/abs/2201.10925} {arXiv:2201.10925 [cond-mat.soft]}
  \BibitemShut {NoStop}%
\bibitem [{Note1()}]{Note1}%
  \BibitemOpen
  \bibinfo {note} {\textcolor {black}{ A higher-order effect, which cannot be
  accounted by Eq.~(\ref {eq:cap-elastica}), it the small torque exerted by the
  liquid-vapor interface on the \textcolor {black}{sheet} if they are not
  perpendicular at the contact line (\protect \textit {i.e.} $\theta _Y
  \protect \neq \pi /2$). This localized torque is explicitly proportional to
  the thickness, yielding a discontinuity of the curvature $[[\kappa ]] \sim
  \cos (\theta _Y) \cdot t/\ell _{bc}^2$, whose effect on the shape is
  negligible , \protect \textit {i.e.} $[[\kappa ']] \gg [[\kappa ]]/\ell
  _{bc}$ for $t \protect \!\ll \protect \! \ell _{bc}$.}}\BibitemShut {Stop}%
\bibitem [{\citenamefont {Schroll}\ \emph {et~al.}(2013)\citenamefont
  {Schroll}, \citenamefont {Adda-Bedia}, \citenamefont {Cerda}, \citenamefont
  {Huang}, \citenamefont {Menon}, \citenamefont {Russell}, \citenamefont
  {Toga}, \citenamefont {Vella},\ and\ \citenamefont
  {Davidovitch}}]{Schroll2013}%
  \BibitemOpen
  \bibfield  {author} {\bibinfo {author} {\bibfnamefont {R.}~\bibnamefont
  {Schroll}}, \bibinfo {author} {\bibfnamefont {M.}~\bibnamefont {Adda-Bedia}},
  \bibinfo {author} {\bibfnamefont {E.}~\bibnamefont {Cerda}}, \bibinfo
  {author} {\bibfnamefont {J.}~\bibnamefont {Huang}}, \bibinfo {author}
  {\bibfnamefont {N.}~\bibnamefont {Menon}}, \bibinfo {author} {\bibfnamefont
  {T.}~\bibnamefont {Russell}}, \bibinfo {author} {\bibfnamefont
  {K.}~\bibnamefont {Toga}}, \bibinfo {author} {\bibfnamefont {D.}~\bibnamefont
  {Vella}},\ and\ \bibinfo {author} {\bibfnamefont {B.}~\bibnamefont
  {Davidovitch}},\ }\bibfield  {title} {\bibinfo {title} {Capillary
  deformations of bendable films},\ }\href@noop {} {\bibfield  {journal}
  {\bibinfo  {journal} {Physical review letters}\ }\textbf {\bibinfo {volume}
  {111}},\ \bibinfo {pages} {014301} (\bibinfo {year} {2013})}\BibitemShut
  {NoStop}%
\bibitem [{\citenamefont {Davidovitch}\ and\ \citenamefont
  {Vella}(2018)}]{Davidovitch2018}%
  \BibitemOpen
  \bibfield  {author} {\bibinfo {author} {\bibfnamefont {B.}~\bibnamefont
  {Davidovitch}}\ and\ \bibinfo {author} {\bibfnamefont {D.}~\bibnamefont
  {Vella}},\ }\bibfield  {title} {\bibinfo {title} {Partial wetting of thin
  solid sheets under tension},\ }\href@noop {} {\bibfield  {journal} {\bibinfo
  {journal} {Soft Matter}\ }\textbf {\bibinfo {volume} {14}},\ \bibinfo {pages}
  {4913} (\bibinfo {year} {2018})}\BibitemShut {NoStop}%
\bibitem [{\citenamefont {Huang}\ \emph {et~al.}(2007)\citenamefont {Huang},
  \citenamefont {Juszkiewicz}, \citenamefont {De~Jeu}, \citenamefont {Cerda},
  \citenamefont {Emrick}, \citenamefont {Menon},\ and\ \citenamefont
  {Russell}}]{Huang2007}%
  \BibitemOpen
  \bibfield  {author} {\bibinfo {author} {\bibfnamefont {J.}~\bibnamefont
  {Huang}}, \bibinfo {author} {\bibfnamefont {M.}~\bibnamefont {Juszkiewicz}},
  \bibinfo {author} {\bibfnamefont {W.~H.}\ \bibnamefont {De~Jeu}}, \bibinfo
  {author} {\bibfnamefont {E.}~\bibnamefont {Cerda}}, \bibinfo {author}
  {\bibfnamefont {T.}~\bibnamefont {Emrick}}, \bibinfo {author} {\bibfnamefont
  {N.}~\bibnamefont {Menon}},\ and\ \bibinfo {author} {\bibfnamefont {T.~P.}\
  \bibnamefont {Russell}},\ }\bibfield  {title} {\bibinfo {title} {Capillary
  wrinkling of floating thin polymer films},\ }\href@noop {} {\bibfield
  {journal} {\bibinfo  {journal} {Science}\ }\textbf {\bibinfo {volume}
  {317}},\ \bibinfo {pages} {650} (\bibinfo {year} {2007})}\BibitemShut
  {NoStop}%
\bibitem [{\citenamefont {Davidovitch}\ \emph {et~al.}(2011)\citenamefont
  {Davidovitch}, \citenamefont {Schroll}, \citenamefont {Vella}, \citenamefont
  {Adda-Bedia},\ and\ \citenamefont {Cerda}}]{Davidovitch2011}%
  \BibitemOpen
  \bibfield  {author} {\bibinfo {author} {\bibfnamefont {B.}~\bibnamefont
  {Davidovitch}}, \bibinfo {author} {\bibfnamefont {R.~D.}\ \bibnamefont
  {Schroll}}, \bibinfo {author} {\bibfnamefont {D.}~\bibnamefont {Vella}},
  \bibinfo {author} {\bibfnamefont {M.}~\bibnamefont {Adda-Bedia}},\ and\
  \bibinfo {author} {\bibfnamefont {E.~A.}\ \bibnamefont {Cerda}},\ }\bibfield
  {title} {\bibinfo {title} {Prototypical model for tensional wrinkling in thin
  sheets},\ }\href@noop {} {\bibfield  {journal} {\bibinfo  {journal}
  {Proceedings of the National Academy of Sciences}\ }\textbf {\bibinfo
  {volume} {108}},\ \bibinfo {pages} {18227} (\bibinfo {year}
  {2011})}\BibitemShut {NoStop}%
\bibitem [{\citenamefont {Schulman}\ and\ \citenamefont
  {Dalnoki-Veress}(2015)}]{Schulman2015}%
  \BibitemOpen
  \bibfield  {author} {\bibinfo {author} {\bibfnamefont {R.~D.}\ \bibnamefont
  {Schulman}}\ and\ \bibinfo {author} {\bibfnamefont {K.}~\bibnamefont
  {Dalnoki-Veress}},\ }\bibfield  {title} {\bibinfo {title} {Liquid droplets on
  a highly deformable membrane},\ }\href@noop {} {\bibfield  {journal}
  {\bibinfo  {journal} {Physical review letters}\ }\textbf {\bibinfo {volume}
  {115}},\ \bibinfo {pages} {206101} (\bibinfo {year} {2015})}\BibitemShut
  {NoStop}%
\bibitem [{\citenamefont {Twohig}\ \emph {et~al.}(2018)\citenamefont {Twohig},
  \citenamefont {May},\ and\ \citenamefont {Croll}}]{Twohig2018}%
  \BibitemOpen
  \bibfield  {author} {\bibinfo {author} {\bibfnamefont {T.}~\bibnamefont
  {Twohig}}, \bibinfo {author} {\bibfnamefont {S.}~\bibnamefont {May}},\ and\
  \bibinfo {author} {\bibfnamefont {A.~B.}\ \bibnamefont {Croll}},\ }\bibfield
  {title} {\bibinfo {title} {{Microscopic details of a fluid/thin film triple
  line}},\ }\href {https://doi.org/10.1039/c8sm01117f} {\bibfield  {journal}
  {\bibinfo  {journal} {Soft Matter}\ }\textbf {\bibinfo {volume} {14}},\
  \bibinfo {pages} {7492} (\bibinfo {year} {2018})},\ \Eprint
  {https://arxiv.org/abs/1804.07797} {arXiv:1804.07797} \BibitemShut {NoStop}%
\bibitem [{\citenamefont {Wagner}\ and\ \citenamefont
  {Vella}(2011)}]{Wagner2011}%
  \BibitemOpen
  \bibfield  {author} {\bibinfo {author} {\bibfnamefont {T.~J.}\ \bibnamefont
  {Wagner}}\ and\ \bibinfo {author} {\bibfnamefont {D.}~\bibnamefont {Vella}},\
  }\bibfield  {title} {\bibinfo {title} {{Floating carpets and the delamination
  of elastic sheets}},\ }\href {https://doi.org/10.1103/PhysRevLett.107.044301}
  {\bibfield  {journal} {\bibinfo  {journal} {Physical Review Letters}\
  }\textbf {\bibinfo {volume} {107}},\ \bibinfo {pages} {1} (\bibinfo {year}
  {2011})}\BibitemShut {NoStop}%
\end{thebibliography}%

\end{document}


\newcommand{\cred}[1]{\textcolor{black}{#1}}
\newcommand{\cblue}[1]{\textcolor{black}{#1}}
\newcommand{\bs}{\bar{s}}
\newcommand{\bkappa}{\bar{\kappa}}
\renewcommand{\theequation}{S\arabic{equation}}

\title{Supplementary Information:\\Peeling from a liquid}

\author{Deepak Kumar$^1$}
\author{Nuoya Zhou${^2}$}
\author{\cblue{Fabian Brau$^{3}$}}
\author{Narayanan Menon${^2}$}%
\author{Benny Davidovitch${^2}$}%
\affiliation{%
 $^1$ Department of Physics, Indian Insitute of Technology Delhi, New Delhi 110016; \\ 
$^2$ Department of Physics, University of Massachusetts Amherst, Amherst MA 01003; \\ 
 \cblue{$^3$ Nonlinear Physical Chemistry Unit, Universite libre de Bruxelles (ULB), CP231, 1050 Brussels, Belgium}}



\maketitle
\section{Supplementary Information}
\subsection{Additional data for the translationally symmetric geometry}
The data presented in Fig. 2 C and D of the main text correspond to situation where the meniscus has been allowed to relax to near-equilibrium configuration. However, when the film is pulled out from the liquid surface or allowed to absorb on to it at small velocity, one obtains two different values of $\psi_0$ corresponding to advancing and receding configurations. $\kappa(s)$ curve in situations close to the advancing and receding angles for PS film of thickness $t=2$ $\mu$m is shown in Fig. \ref{fig:S2} A and Fig. \ref{fig:S2} B respectively. The cusp in $\kappa(s)$ at the three phase contact is clearly seen even in these situations with different values of $\psi_0$.

\subsection{Examining the data analysis protocol}
\subsubsection{Case I: Smooth curvature}
In order to make sure that no spurious artifact is introduced by the data analysis algorithm adopted, we take a parabolic function $y=0.7x^2$ (Fig. \ref{fig:S1} A) which has a shape (in length units of mm) similar near the origin to the typically measured experimental profile. The function is sampled to generate $(x,y)$ coordinates. The data set so generated is run through the same algorithm used in the main text to obtain $\kappa(s)$, plotted in Fig. \ref{fig:S1}B as filled black circles. We have used $\ell_{bc}=0.2$ mm, a value close to that in the experiments. The exact analytical result obtained for the particular function chosen is plotted as a red solid line. The two match exactly with each other in this case - a parabola has a smooth $\kappa(s)$ curve which the data analysis algorithm captures correctly without introducing any spurious cusp at the origin. 

\subsubsection{Case II: Curvature with a cusp}
We consider the following two polynomials:
\begin{equation}
    f_1(x)=-0.023x^4-0.25x^3+0.72x^2+0.0023x-0.0055
\end{equation}
\begin{equation}
    f_2(x)=0.088x^4+0.38x^3+0.72x^2+0.0023x-0.0055
\end{equation}

Using the above two polynomials we define the following piece-wise function:
\begin{equation} 
y(x)= \left\{
  \begin{array}{lr} 
      f_1(x) & x< 0 \\
      f_2(x) & x > 0 
      \end{array}
\right.
\label{cuspy_function}
\end{equation}
The coefficients of the polynomials have been chosen to keep the shape (in length units of mm) of the function $y(x)$ similar to a typical experimentally measured profile. The functions have been plotted in Fig. \ref{fig:S3} A (we have assumed $\ell_{bc}=0.2$ mm). It may be noted that the two polynomials have the same coefficients for the constant, linear and square terms but different coefficients for the cubic and the quartic terms. This implies that at $x=0$, the function $y(x)$ defined in Eq. \ref{cuspy_function} has a continuous but non-smooth curvature. The expected $\kappa(s)$ curve is shown as solid lines in Fig. \ref{fig:S3} B .The function $y(x)$ in Eq. \ref{cuspy_function} is used to generate data points $(x,y)$ at a sampling rate comparable to the experiments. When the data analysis algorithm is employed on the polynomial functions $f_1(x)$ and $f_2(x)$, the curvature (open circles in Fig. \ref{fig:S3} B) match exactly with the expected result.

Curvature was obtained by using our data analysis algorithm on the data generated by sampling the function $y(x)$ is shown in Fig. \ref{fig:S4}A. The red, blue and green curves show the curvature computed by taking $\Delta=\ell_{bc}$, $0.5\ell_{bc}$ and $0.2\ell_{bc}$ respectively while the black curve shows the analytically calculated result. We next study the effect of noise present in experimentally measured data. In panels B and C of Fig. \ref{fig:S4} a uniformly distributed noise of width corresponding to 1 and 5 pixels, respectively, have been added to the function $y(x)$ before sampling it to obtain the data for analysis. We observe that a smaller value of $\Delta$ does a better job of reproducing the expected curvature near the vertex in absence of noise. However, in the presence of noise, a larger value of $\Delta$ is needed to average out the effect of noise. The value $\Delta=\ell_{bc}$ chosen for the analysis is an optimum compromise for the typical level of noise present in the experimental data.  

As described in the main text, in our data analysis algorithm the data is sampled in a different way from the interval containing the vertex than from the other intervals. In each iteration, we choose one data point randomly with uniform probability from all other intervals, however, from the interval containing the vertex the data point is chosen randomly with a normal probability distribution centered at position of vertex. Fig. \ref{fig:S5} shows the significance of this step. If we sample the interval containing the vertex in the same way as other intervals, the position of the cusp varies slightly from iteration to iteration. As a result when the average is calculated the cusp is smoothed out (Fig. \ref{fig:S5} blue curve). On the other hand choosing the vertex preferentially reduces this noise and helps in capturing the non-smooth nature of the curvature (Fig. \ref{fig:S5} red curve).

\subsection{Experimental details regarding the axially symmetric case}

In the axially symmetric case, an air bubble is introduced underneath a thin film floating on water surface to create a three-phase contact line. In order to image the film profile, the polystyrene film is flourescently tagged with the dye Nile Red. Before spin-coating polystyrene, Nile Red is dissolved in polystyrene-toluene solution at a concentration of $10^{-6}$wt $\%$. After spin coating, we obtain films that can be excited by a sheet of green laser, using a scheme similar to that illustrated in Fig. 2A of the main text, to get the solid-vapor interface. The red-channel of a typical image obtained using this method is shown in Fig.\ref{fig:S6} A. As can be noticed in the image, the liquid-vapor interface is not visible in this scheme. We, therefore, also capture a bright field image using the same setup under normal room lighting, which reveals the lower part of the air-bubble as shown in Fig. \ref{fig:S6} B. In Fig. \ref{fig:S6} C, we show an image where Fig. \ref{fig:S6} B has been overlayed on Fig. \ref{fig:S6} A. The red dashed curve shows a circular curve fitted to the visible part of the air bubble.

\subsection{Note on error bars in the measurements}

In order to compute the curvature, we divide the whole data set into intervals of length $\Delta$ and choose one data point from each interval. A spline function of order $3$ made up of Hermite polynomials is generated using the data points sampled in each iteration ($i$) and the curvature $\kappa_i (s)$ is computed on this spline function at every $n^{th} (\sim 10)$ point of the original data. The process is repeated a large number of times ($N$), selecting a different sampling of data points randomly in each iteration, such that the whole data set is adequately represented. The curvature profiles obtained from individual data samples are averaged to obtain the final $\kappa(s)$ curve: $\kappa(s)=\frac{1}{N}\sum_{i=1}^{N}\kappa_i(s)$. The variance in $\kappa(s)$ is computed as $\sigma_{\kappa}^{2}(s)=\frac{1}{N}\sum_{i=1}^N (\kappa_i(s)-\kappa(s))^2$. The error bars on $\kappa(s)$ are computed as $\sigma_{\kappa}(s)/\sqrt{n}$. Fig. 2C of the main text shows the typical error bars measured.

\makeatletter 
\renewcommand{\thefigure}{S\@arabic\c@figure}
\makeatother

\clearpage
\begin{figure}
    \includegraphics[width=0.8\textwidth]{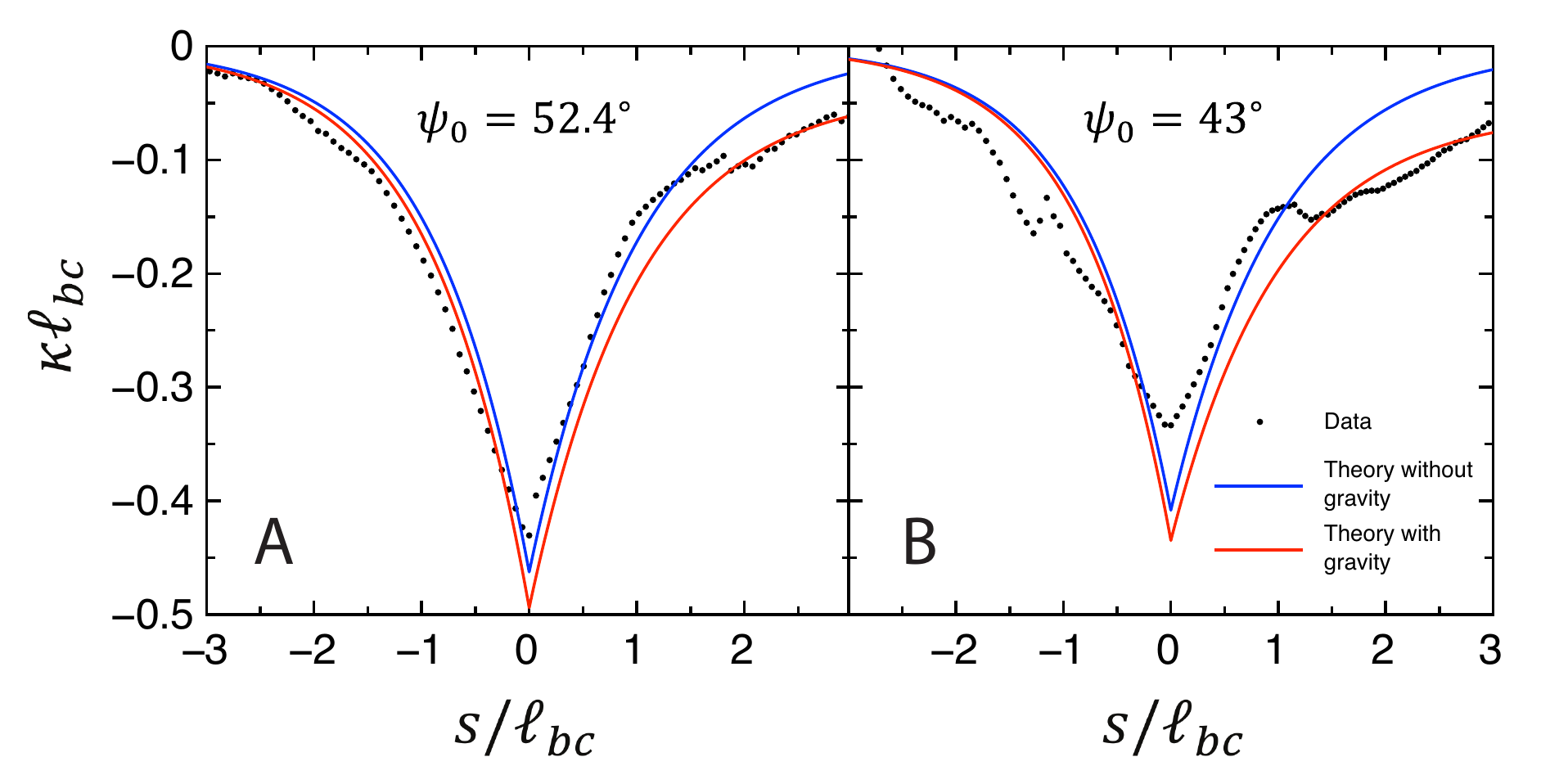}
    \caption{$\kappa(s)$ for polystyrene film of thickness $t=2$  $\mu$m at (A)  $\psi_0=52.4^\circ$ close to the advancing configuration and (B) at $\psi_0=43^\circ$ close to the receding configuration. The cusp in curvature near the vertex is clearly seen.}
    \label{fig:S2}
\end{figure}

\clearpage
\begin{figure}
    \includegraphics[width=0.8\textwidth]{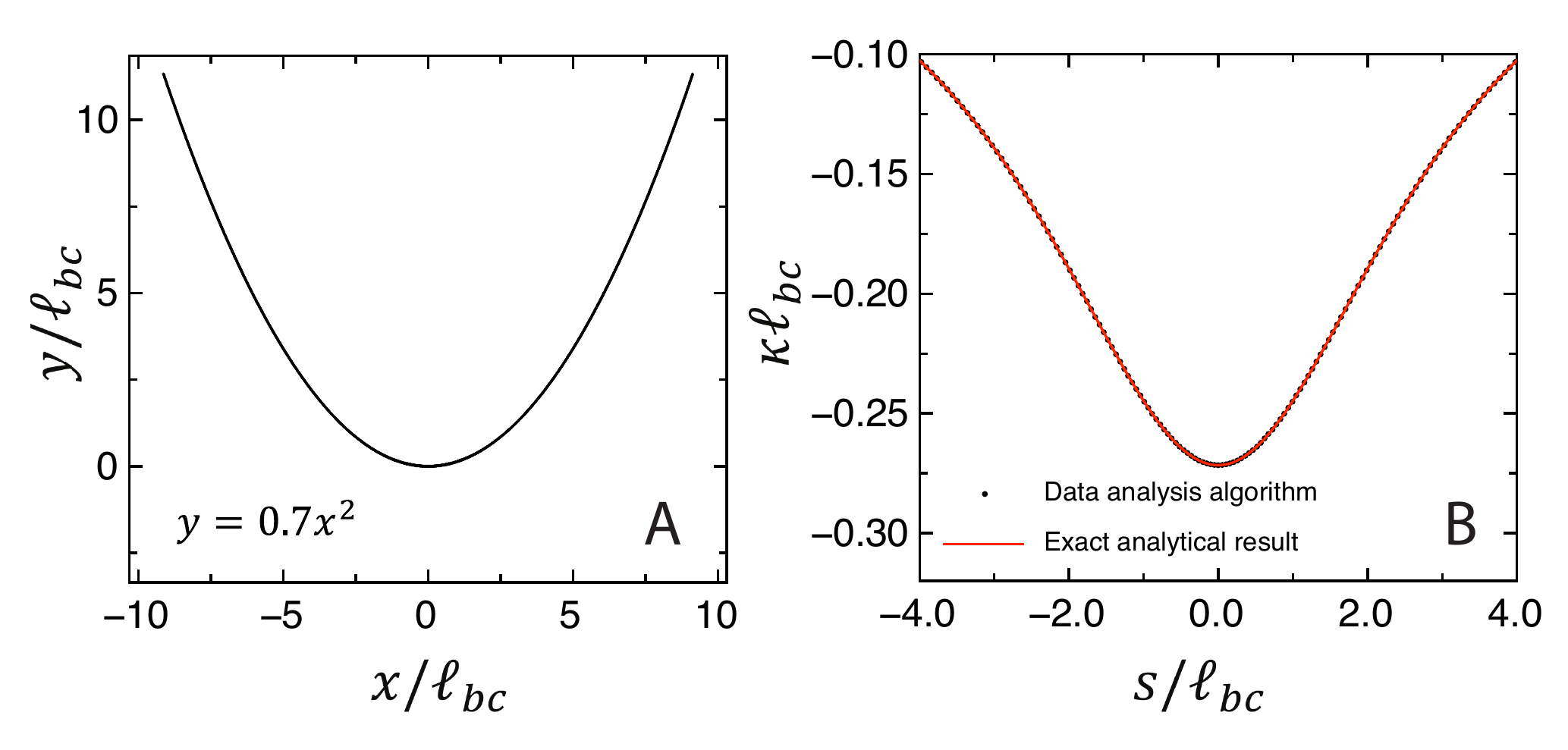}
    \caption{(A) Data sampled from the function $y=0.7 x^2$. (B) Black circles: Curvature $\kappa(s)$ computed from the data in (A) using the algorithm used in the paper. Red solid curve: $\kappa(s)$ obtained from the function $y=0.7x^2$ analytically. The two match exactly.}
    \label{fig:S1}
\end{figure}

\clearpage
\begin{figure}
    \includegraphics[width=0.8\textwidth]{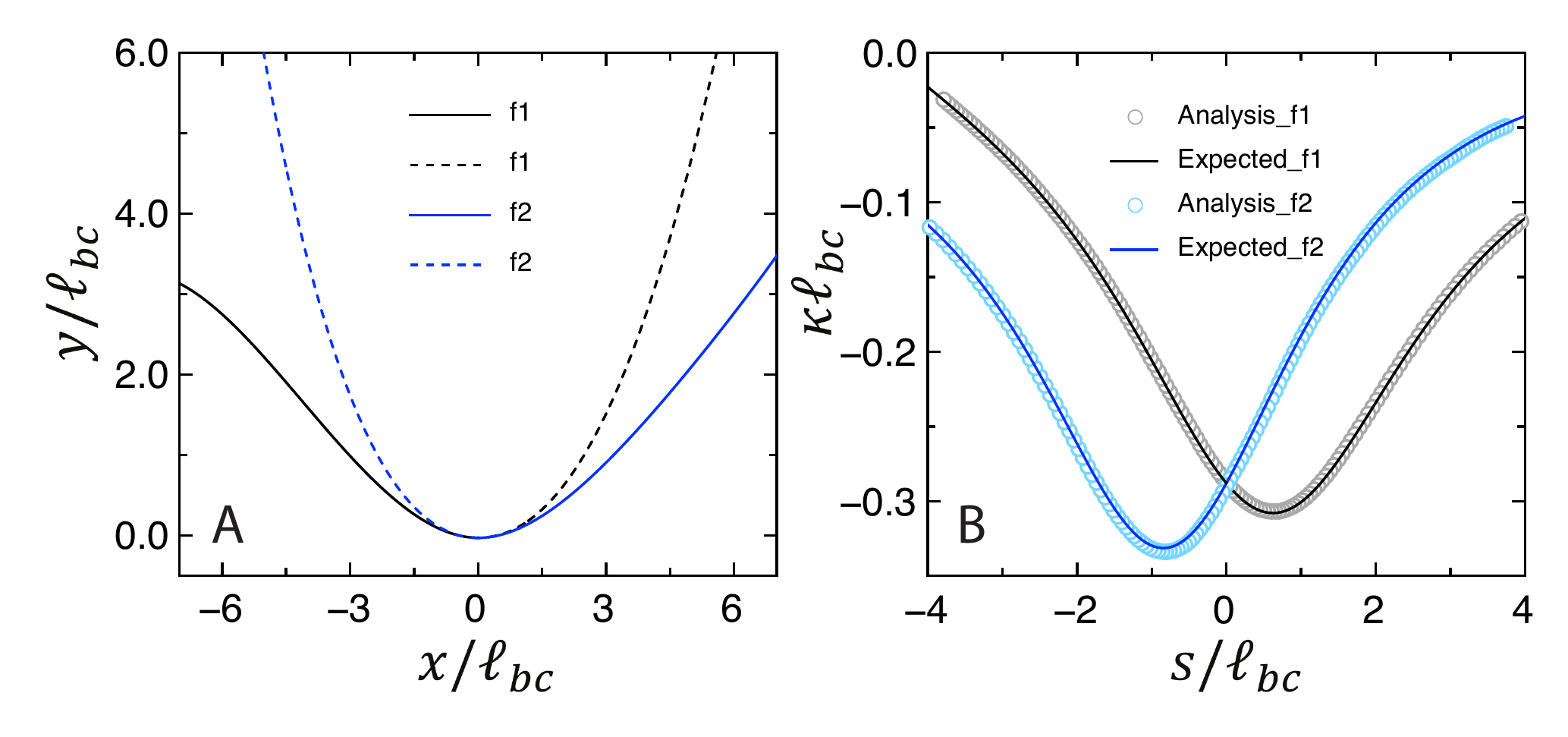}
    \caption{(A) Two polynomial functions $f_1(x)$ and $f_2(x)$ are used to define a piecewise function $y(x)$ shown as solid line. (B) $k(s)$ curve obtained for the functions $f_1(x)$ and $f_2(x)$ using the data analysis algorithm (open circles) match exactly with the expected result (solid lines).}
    \label{fig:S3}
\end{figure}

\clearpage
\begin{figure}
    \includegraphics[width=1\textwidth]{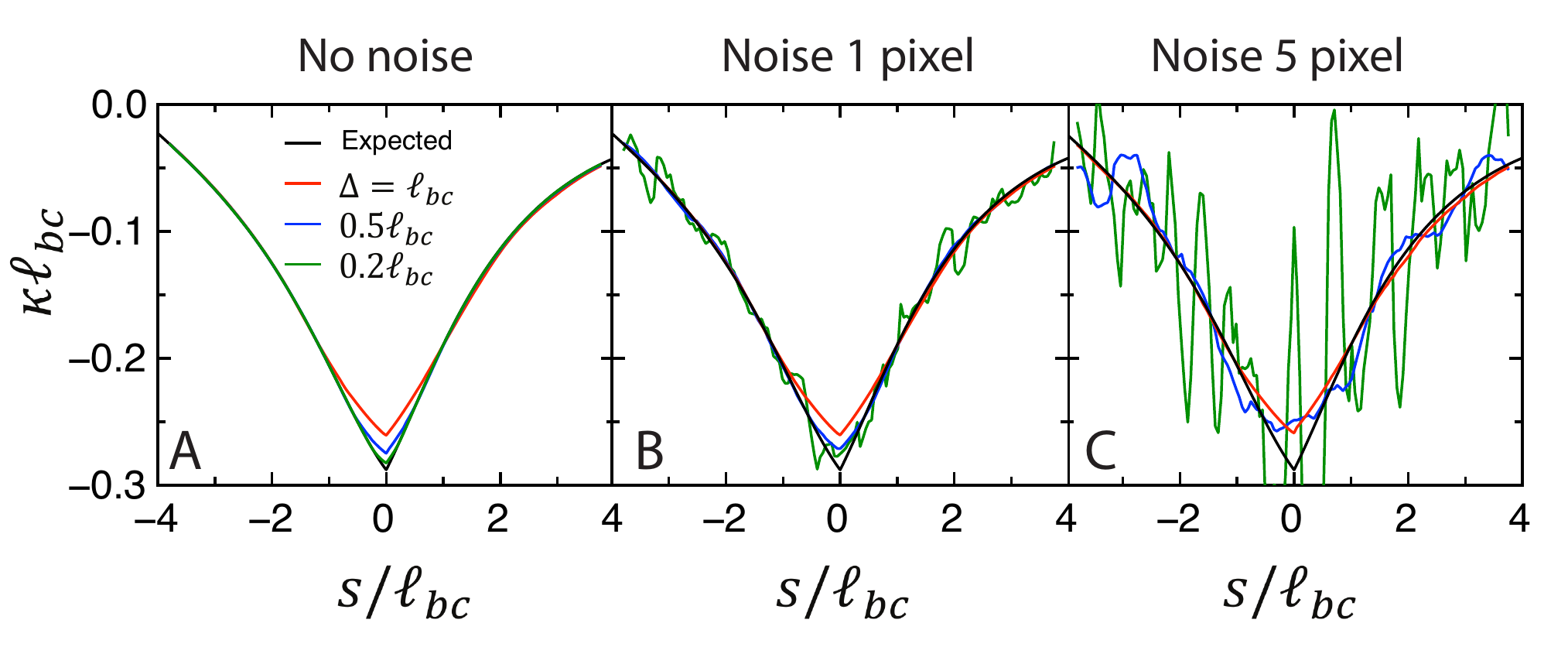}
    \caption{Effect of varying the value of $\Delta$ in the data analysis protocol. Smaller values of $\Delta$ give a better match to the expected curvature in the limit of low noise. As the noise in data is increased the curves obtained with lower values of $\Delta$ become more noisy. $\Delta\approx\ell_{bc}$ seems to be an optimal choice for $\Delta$.}
    \label{fig:S4}
\end{figure}

\clearpage
\begin{figure}
    \includegraphics[width=0.7\textwidth]{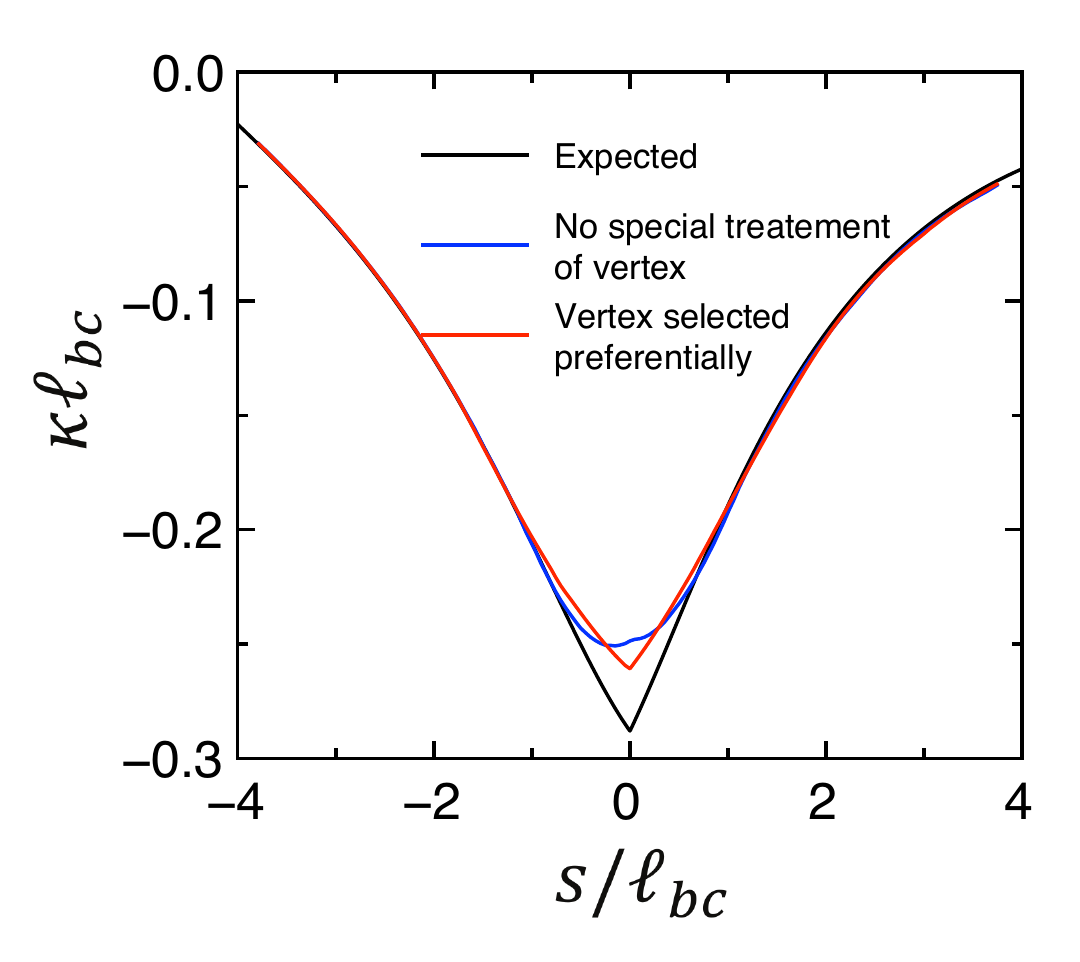}
    \caption{The blue curve shows the curvature $\kappa(s)$ obtained if we allow each data point in the interval containing the vertex to be chosen with equal probability. We notice that the cusp in the curvature present in the actual function (black curve) is smoothed out. In the data analysis protocol adopted, in the interval containing the vertex, we select points with a normal probability distribution centered at the vertex and having a width comparable to the typical error in the measurement of the vertex (red curve). }
    \label{fig:S5}
\end{figure}

\clearpage
\begin{figure}
    \includegraphics[width=1\textwidth]{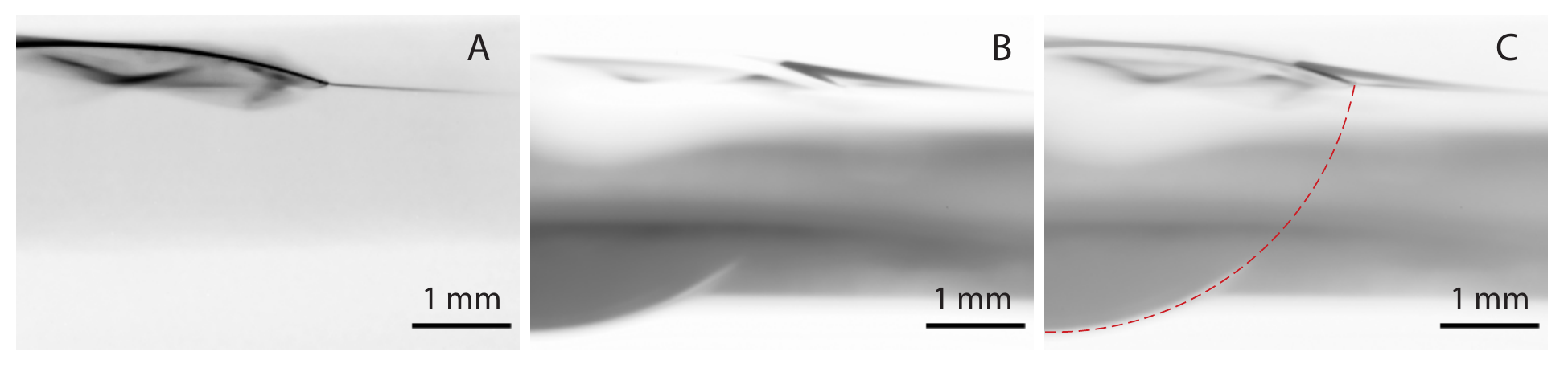}
    \caption{(A) Image (red-channel) of the film profile in the axially-symmetric geometry obtained using fluorescent imaging. (B) Bright field image capturing a part of the air-bubble. (C) Image B overlayed on inage A. The red dashed curve is a circle fitted to the visible part of the air-bubble. }
    \label{fig:S6}
\end{figure}

\section{Theory}
Here we elaborate on the solution of the capillary \textit{elastica}, Eq. (9), which is given by Eqs.~(11,12) and shown in the solid blue curves in Fig.~2B-D (and the correction due to the effect of liquid gravity, shown in red solid curves in Fig.~2C-D and dashed blue in Fig.~2B).  

On both sides of the contact line, the profile of the film is determined by Eq.~(9). If liquid gravity is ignored, in each of the the regions  ($s>0$ and $s<0$) Eq.~(9) can be considered a nonlinear $3^{rd}$    
equation (ODE) for the angle $\psi(s)$, whose solution requires 3 boundary conditions (BCs). 
Eqs.~(7) and (10)
provide two of those six BCs. 
In order to specify the other four BCs, we use the normalized arclength and curvature as defined in the main text:  
$\bs = s/\ell_{bc}$, $\bkappa = \kappa \cdot \ell_{bc}$, and consider the BCs at $\bs \to \pm \infty$. In the terminology of singular perturbation theory, we seek to provide boundary conditions for Eq.~(9)
in the ``inner region'', $|\bs| \sim O(1)$, by matching to an ``outer region'', 
$|\bs| \gg 1$. In the dry part, the film reaches a straight vertical profile, such that: $\psi, \bkappa 
\to 0$ as $\bs \to -\infty$. In the wet part, we consider distances $\ell_{bc} \ll s \ll \ell_c$ from the contact line, such that $\bs \sim O(1/\sqrt{\epsilon}) \to \infty$, where we recall that $\epsilon \ll 1$. In this ``intermediate'' region, the bending-dominated vicinity of the contact line appears as a point (the ``vertex''), whereas the curvature due to the the gravity-dominated meniscus is small 
($\sim O(\ell_c^{-1})$), 
such that the surface appears as a slightly curved plane,  inclined at an angle $\approx \psi_{0}$ to the vertical. 
The normalized curvature, $\bkappa \approx \sqrt{\epsilon} f(\psi_0)$,  
where $\psi_0$ and $f(\psi_0)$ are given by Eq.~(7), 
such that at $O(\epsilon^0)$ the six BCs are: 
\begin{eqnarray}
    \bs \to -\infty &:& \ \ \psi \to 0 \ ; \  \bkappa 
    \to 0  \nonumber \\
    \bs \to +\infty &:& \ \psi \to \psi_0 \ ;   
    \ \bkappa 
    \to 0 \nonumber \\
    \bs=0 &:& \ \ [[\bkappa]] = 0 \ , \ [[\bkappa']] = -\sin\theta_Y
    \label{eq:BCs}
\end{eqnarray}
Furthermore, at $O(\epsilon^0)$ the hydrostatic pressure in Eq.~(9) 
is negligible ($P\ell_{bc}^3/B \!\sim\! \sqrt{\epsilon}$), and the equation becomes
integrable. The exact solution, 
which satisfies all BCs~\eqref{eq:BCs} is given by Eqs.~(11, 12). The tangent angle is readily obtained by integrating the curvature, yielding: 
\begin{eqnarray}\  
\bs \! &>& \! 0 \ : \  
\psi \!=\! \psi_0  \!+\! \pi \!-\! 4\tan^{-1} \text{tanh}\frac{\bs + \bs_w}{2}   \nonumber \\ 
\bs \! &<& \! 0 \ : \ 
 \  \psi \!= - \pi - \! 4\tan^{-1} \text{tanh}[\frac{\bs + \bs_d}{2}\sqrt{r}]   \  \label{eq:oreder-0} \end{eqnarray}


To facilitate comparison with experimental data, it is useful to incorporate the effect of liquid gravity in Eq.~(9). When focusing on the ``inner'' region (\textit{i.e.} vicinity $|s| \lesssim \ell_{bc}$ of the contact line), this amounts to computing the $O(\epsilon)$ corrections to the above $O(\epsilon^0)$ solution. This is obtained by substituting the formal expansion: $\bkappa = \bkappa_0 + \sqrt\epsilon \bkappa_1 +\cdots$, in
Eq.~(9) 
where $\bkappa_0$ is given by Eq.~(11),  
and solving the resulting (linear) ODE for $\bkappa_1$ (where the BCs at $\bs \to \infty$ include now the $O(\sqrt{\epsilon})$ terms that were neglected in Eq.~(\ref{eq:BCs})). Even though the relative effect of the correction $\sqrt\epsilon \bkappa_1$ is small in the inner region, where $\bkappa_0$ is finite, it eventually dominates the curvature in the ``outer" region, $\bs \to \infty$, where $\bkappa_0 \to 0$. Specifically, for $\bs \gg 1$, one finds that the dominant term on the LHS of Eq.~(9) is $-\gamma_{lv} \kappa $, whereas the RHS is approximated by $-P \approx \rho g \ell_c f(\psi_0) \approx \sqrt{ \rho g \gamma_{lv}} f(\psi_0) $ (where $f(\psi_0)$ is given by Eq.~(7)). Consequently, at $\bs \to \infty$ we obtain: 
$\bkappa \approx \sqrt{\epsilon} \bkappa_1 \approx - \sqrt{\epsilon} f(\psi_0)$. This small, finite asymptotic value of the curvature in the wet part of the sheet is readily observed in the red curves in Fig.~2C,2D, and its effect is noticeable upon comparing the solid and dashed blue curves in Fig.~2B.

\bibliography{Ref}